\documentclass[12pt]{article}
\begin{document}

\addtolength{\baselineskip}{0.5\baselineskip}

\title{\textbf{Orbital Approximation for the Reduced Bloch Equations: Fermi-Dirac
Distribution for Interacting Fermions and Hartree-Fock Equation at
Finite Temperature}}
\author{Liqiang Wei\\
Institute for Theoretical Atomic, Molecular and Optical Physics\\
Harvard University, Cambridge, MA 02318\\
%Department of Chemistry, University of Illinois \\
%Urbana, IL 61801\\
\\
Chiachung Sun\\
%State Key Laboratory of Theoretical and Computational Chemistry\\
%Jilin University,
Institute of Theoretical Chemistry, Jilin University\\
Changchun, Jilin 130023 P. R. China}

\maketitle

\begin{abstract}
\vspace{0.05in} In this paper, we solve a set of hierarchy
equations for the reduced statistical density operator in a grand
canonical ensemble for an identical many-body $\it{fermion}$
system without or with two-body interaction. We take the
single-particle approximation, and obtain an eigen-equation for
the single-particle states. For the case of no interaction, it is
an eigen-equation for the free particles, and solutions are
therefore the plane waves. For the case with two-body interaction,
however, it is an equation which is the extension of usual
Hartree-Fock equation at zero temperature to the case of any
finite temperature. The average occupation number for the
single-particle states with mean field interaction is also
obtained, which has the same Fermi-Dirac distribution form as that
for the free fermion gas. The derivation demonstrates that even
for an interacting fermion system, only the lowest $N$ orbitals,
where $N$ is the number of particles, are occupied at zero
temperature. In addition, their practical applications in such
fields as studying the temperature effects on the average
structure and electronic spectra for macromolecules are discussed.
\\
\\
$\underline{PACS}$ 05.30.-d; 31.70.-f; 71.10.-w; 72.80.Le
\end{abstract}

\vspace{0.35in}
\section{Introduction}

Quantum statistical mechanics provides a most general theoretical
framework for studying the structure and dynamics of an
interacting many-body system. Combined with mathematical laws of
statistical distribution, it investigates the microscopic quantum
states of a system with many degrees of freedom, its corresponding
macroscopic thermodynamic behaviors, and their interplay [1-4]. It
covers the usual quantum mechanics as a special case at zero
temperature [5].

Experimental investigation of temperature or pressure effects on
the electronic structure and related spectra for molecules or
solids has been for some time [6-8]. However, a complete or
systematic theoretical work has not been developed and seen in the
literature. The quantum chemistry at finite temperature or
pressure is not a well-defined or well-established field [9].
Indeed, the effect of temperature or pressure on microscopic
structure is a complicated phenomenon. There exists different
functioning mechanisms. One consideration is that the variation of
the temperature, according to the Fermi-Dirac statistics, will
change the thermal probability distribution of single-particle
states for a free electron gas. Similar situation should be
expected to occur for an interacting electron system, and
therefore its microscopic structure will be correspondingly
altered. Another consideration is that, for molecules or solids,
the thermal excitation will cause the change of the time scales
for the molecular motions. This will most likely bring about the
transitions of electronic states, and therefore lead to the
breakdown of the Born-Oppenheimer approximation. Electron-phonon
interaction is a fundamental topic in solid state physics and its
temperature dependence is well-known. In this article, we tackle
the issues in a simpler way. We treat only an interacting
identical $\it{fermion}$ system, or neglect the coupling of the
electronic motion with those of the nucleus in molecules or
solids. We expect that some sort of the general conclusions will
come out from this study. As a matter of fact, this is also the
approach usually adopted in non-adiabatic molecular dynamics, in
which purely solving the eigenequation for the electrons will
provide the reference states for investigating the coupling
motions between the electrons and the nucleus of the molecules.

In a recent paper [10], we have deduced a set of hierarchy
equations for the reduced statistical density operators in both
canonical and grand canonical ensembles. They provide a law
according to which the reduced density matrix varies in
temperature. They also offer a route for a direct determination of
the reduced density operator for the statistical ensembles.

This paper is an initial endeavor, on the basis of the equations
we derived in paper [10], to investigate the issues related to the
interplay between the microscopic structure and macroscopic
observables, and specifically, the temperature effects on the
electronic structure of molecules or solids. We take the orbital
model and solve the equations for the grand canonical ensemble in
an approximate way. In the next Section, we first consider the
case of a free electron gas, and evaluate the terms related to the
single-particle operator of the reduced Bloch equations, we get
the solutions of a plane wave as well as the usual Fermi-Dirac
statistics of a free electron gas. In Section 3, we solve the
equations for the identical fermion system with two-body
interaction. We obtain an energy equation for the single-particle
states which is the extension of usual Hartree-Fock equation at
zero temperature to the case of any finite temperature. The
average occupation number is also obtained which has the same form
as that for a free fermion gas, which is the Fermi-Dirac
distribution. The Section 4 is a summary and conclusion. The
implications and possible applications are discussed.

\vspace{0.35in}
\section{Free Fermion Gas}

In this and next sections, we consider the solution of reduced
Bloch equations for the grand canonical ensemble we derived in
paper [10] as shown below,
\begin{eqnarray}
\nonumber
-\frac{\partial}{\partial\beta}D^{p}&=&\bar{H}^{p}_{1}D^{p}+(p+1)L^{p}_{p+1}
\left[\bar{h}(p+1)D^{p+1}\right]+(p+1)L_{p+1}^{p}\left[\sum_{i=1}^{p}g(i,p+1)D^{p+1}\right]+\\
&&+\left(\begin{array}{c} p+2\\2\end{array}\right)
L_{p+2}^{p}\left[g(p+1,p+2)D^{p+2}\right],
\end{eqnarray}
where
\begin{equation}
  \bar{H}_{1}^{p} = \sum_{i=1}^{p}\bar{h}(i) + \sum_{i<j}^{p}
  g(i,j),
\end{equation}
and
\begin{equation}
 \bar{h}(i) = h(i)-\mu.
\end{equation}
The $h(i)$ and $g(i,j)$ are one- and two-body operators, and the
rest of symbols are defined in [10]. We choose a grand canonical
ensemble for our study because we want to compare our results with
those already known for a free fermion gas. Just like solving any
other equations of motion, the exact solutions can be obtained
only for very few cases. Some sort of approximations have to be
made. Nevertheless, more universal conclusions might be possibly
obtained. A common approximation scheme is to truncate the
hierarchy. Depending on the specific physical situation, this is
accomplished by approximating the higher-order density matrix as a
functional of the lower-order ones. In this paper, the system
considered is $N$-fermions and the decoupling scheme we assume is
to express $D^{p+1}$ by $D^{p}$ in the following way [11-15],
\begin{eqnarray}
\nonumber D^{p+1}&=&D^{p}\wedge D^{1}/D^{0} \\
\nonumber &=& \underbrace{D^{1}\wedge D^{1} \wedge ... \wedge
D^{1}}_{p+1}/(D^{0})^{p},
\end{eqnarray}
and
\begin{eqnarray}
\nonumber D^{p+2}&=&D^{p+1}\wedge D^{1}/D^{0} \\
&=& \underbrace{D^{1}\wedge D^{1} \wedge ... \wedge
D^{1}}_{p+2}/(D^{0})^{p+1}.
\end{eqnarray}
That is, the $p$th-order reduced density matrix can be expressed
as a $p$-fold Grassmann product of its first-order reduced density
matrices. This implies that we have taken an independent-particle
approximation. To proceed further, we need to evaluate various
terms of right hand side of Eq. (1).

We first study the case without any interaction, where we only
need to calculate the first and second terms of Eq. (1). For the
first term,
\begin{eqnarray}
\nonumber \bar{H}^{p}_{1}\cdot D^{p}&=&\bar{H}^{p}_{1}\cdot
\underbrace{D^{1}\wedge D^{1} \wedge ... \wedge D^{1}}_{p}/(D^{0})^{p-1} \\
&=& \sum_{i=1}^{p}\underbrace{D^{1}\wedge D^{1} \wedge ... \wedge
\bar{h}_{(i)}\cdot D^{1}\wedge ...\wedge D^{1}}_{p}/(D^{0})^{p-1}.
\end{eqnarray}
For the second term, the calculation is more complex, and the
following formulae related to the decomposition of a permutation
group in terms of its subgroups are needed [16],
\begin{eqnarray}
\nonumber A_{p+1} &=& \frac{1}{p+1}\left[A_{p}-\sum_{i=1}^{p}A_{p}\cdot
(p+1,i)\right] \\
&=&\frac{1}{p+1}\left[A_{p}-\sum_{i=1}^{p}(p+1,i)\cdot
A_{p}\right],
\end{eqnarray}
and
\begin{equation}
 \bar{h}(p+1)\cdot A_{p}\cdot(p+1,i)=
 A_{p}\cdot(p+1,i)\cdot]\bar{h}(i),
\end{equation}
where $A_{p}$ is a $p$th-order antisymmetric operator, and
$(p+1,i)$ is an exchange between $p+1$ and $i$. Therefore,
\begin{eqnarray}
\nonumber &&(p+1)L_{p+1}^{p}\left[\bar{h}(p+1)D^{p+1}\right] \\
\nonumber
&=&(p+1)L_{p+1}^{p}\left[\bar{h}(p+1)A_{p+1}\left(\underbrace{D^{1}\otimes
D^{1}\otimes ...\otimes D^{1}}_{p+1}\right)
A_{p+1}\right]/(D^{0})^{p} \\
\nonumber&=&\left(Tr\bar{h}D^{1}\right)\underbrace{D^{1}\wedge
D^{1}\wedge ...\wedge D^{1}}_{p}/(D^{0})^{p}-\\
&&-\sum_{i=1}^{p}\underbrace{D^{1}\wedge...\wedge
D^{1}\bar{h}_{(i)}D^{1}\wedge ...\wedge D^{1}}_{p}/(D^{0})^{p}.
\end{eqnarray}
For the special case of $p=1$, inserting Eqs. (5) and (8) into Eq.
(1) yields
\begin{equation}
-\frac{\partial}{\partial \beta}D^{1} = \bar{h} D^{1}+
\frac{Tr\left(\bar{h}D^{1}\right)}{D^{0}}D^{1}-\frac{1}{D^{0}}D^{1}\bar{h}D^{1}.
\end{equation}
This is the Bloch equation for the first-order reduced density
matrix of an identical particle system with no interaction under
single-particle approximation.

Define
\begin{equation}
  \rho^{1} = D^{1}/D^{0},
\end{equation}
and then Eq. (9) can be simplified as
\begin{equation}
-\frac{\partial}{\partial\beta}\rho^{1}
=\bar{h}\rho^{1}-\rho^{1}\bar{h}\rho^{1}.
\end{equation}
From above equation and its conjugate, we get
\begin{equation}
  h\rho^{1}-\rho^{1}h = 0.
\end{equation}
This means that $h$ and $\rho^{1}$ commute. Since they are also
Hermitian, they have common eigenfunctions $\{|\phi_{i}>\}$, which
are plane waves,
\begin{equation}
  h|\phi_{i}>=\epsilon_{i}|\phi_{i}>,
\end{equation}
and
\begin{equation}
\rho^{1}|\phi_{i}>=\omega(\beta, \mu, \epsilon_{i})|\phi_{i}>
\end{equation}
with
\begin{equation}
  \rho^{1} =
  \sum_{i}\omega(\beta,\mu,\epsilon_{i})|\phi_{i}><\phi_{i}|.
\end{equation}
Substituting Eq. (15) into Eq. (11), we can obtain the equation
the thermal probability $\omega(\beta,\mu,\epsilon_{i})$
satisfies,
\begin{equation}
-\frac{\partial}{\partial\beta} \omega(\beta, \mu, \epsilon_{i})=
(\epsilon_{i}-\mu)\omega(\beta,\mu,\epsilon_{i})-(\epsilon_{i}-\mu)\omega^{2}
(\beta,\mu,\epsilon_{i}).
\end{equation}
Its solution takes the form,
\begin{equation}
\omega(\beta,\mu,\epsilon_{i})=\frac{1}{1+e^{\beta(\epsilon_{i}-\mu)}},
\end{equation}
which is just the usual Fermi-Dirac distribution. Therefore, by
directly solving the reduced Bloch equations for the reduced
density matrix in a grand canonical ensemble under orbital
approximation, we not only have obtained the eigen solutions of a
plane wave but also have recovered the usual Fermi-Dirac
distribution for the free electron gas.

\vspace{0.35in}
\section{Hartree-Fock Equation at Finite Temperature}

In this section, we solve the reduced Bloch equation (1) for the
case of an identical fermion system with two-body interaction
under the orbital approximation (4).The first term and the second
term remain the same as those for the free fermion case. We need
to evaluate the remaining two terms. In particular, for $p=1$, the
Eq. (1) reads
\begin{equation}
-\frac{\partial}{\partial\beta}
D^{1}=\bar{h}D^{1}+\frac{Tr(\bar{h}D^{1})}{D^{0}}D^{1}-\frac{1}{D^{0}}D^{1}\bar{h}
D^{1}+2L^{1}_{2}\left[g(1,2)D^{2}\right]+3L^{1}_{3}\left[g(2,3)D^{3}\right].
\end{equation}
The last two terms can be evaluated in a straightforward way,
which yields
\begin{equation}
 2L^{1}_{2}\left[g(1,2)D^{2}\right] = (J-K)D^{1},
\end{equation}
and
\begin{equation}
 3L^{1}_{3}\left[g(2,3)D^{3}\right]
 =\frac{Tr(gD^{2})}{D^{0}}-\frac{1}{D^{0}}D^{1}(J-K)D^{1},
\end{equation}
where
\begin{equation}
J = Tr_{2}\left[g\cdot D^{1}(2;2)\right]/D^{0},
\end{equation}
and
\begin{equation}
 K = Tr_{2}\left[g\cdot (2,3)\cdot D^{1}(2;2)\right]/D^{0},
\end{equation}
are called the Coulomb and exchange operators, respectively. The
action of $K$ is
\begin{eqnarray}
\nonumber K\cdot D^{1}(3;3)&=& Tr_{2}\left[g\cdot (2,3)\cdot D^{1}(2;2)\right]
/D^{0}\cdot D^{1}(3;3) \\
&=& Tr_{2}\left[g\cdot D^{1}(3;2)\cdot D^{1}(2;3)\right]/D^{0}.
\end{eqnarray}
Substitution of Eqs. (19) and (20) into Eq. (18) yields the Bloch
equation for the first-order reduced density matrix of $N$
interacting fermions under orbital approximation,
\begin{equation}
-\frac{\partial}{\partial\beta}D^{1}=(F-\mu)D^{1}+\left(\frac{Tr\bar{h}D^{1}}{D^{0}}
+\frac{TrD^{2}}{D^{0}}\right)D^{1}-\frac{1}{D^{0}}D^{1}(F-\mu)D^{1},
\end{equation}
where
\begin{equation}
    F = h+J-K,
\end{equation}
is called the Fock operator at finite temperature. Redefine the
normalized first-order reduced density operator
\begin{equation}
 \rho^{1} = D^{1}/D^{0},
\end{equation}
we can simply above equation into
\begin{equation}
-\frac{\partial}{\partial\beta} \rho^{1}
=(F-\mu)\rho^{1}-\rho^{1}(F-\mu)\rho^{1}.
\end{equation}

From Eq. (27) and its conjugate, we get
\begin{equation}
  F\rho^{1}-\rho^{1}F = 0,
\end{equation}
which means that the Fock operator $F$ and $\rho^{1}$ commute.
They are also Hermitian, and therefore they have common
eigenvectors $\{|\phi_{i}>\}$. These vectors are determined by the
following eigen equation for the Fock operator,
\begin{equation}
  F|\phi_{i}>=\epsilon_{i}|\phi_{i}>.
\end{equation}

The first-order reduced density operator is correspondingly
expressed as
%\begin{equation}
%\rho^{1}|\phi_{i}>=\omega(\beta, \mu, \epsilon_{i})|\phi_{i}>
%\end{equation}
%with
\begin{equation}
  \rho^{1} =
  \sum_{i}\omega(\beta,\mu,\epsilon_{i})|\phi_{i}><\phi_{i}|,
\end{equation}
where $\omega(\beta,\mu,\epsilon_{i})$ is the thermal probability
that the orbital is found to be in the state $\{|\phi_{i}>\}$ at
finite temperature $T$. Substituting Eq. (30) into Eq. (27), we
can obtain the equation this thermal probability
$\omega(\beta,\mu,\epsilon_{i})$ satisfies,
\begin{equation}
-\frac{\partial}{\partial\beta} \omega(\beta, \mu, \epsilon_{i})=
(\epsilon_{i}-\mu)\omega(\beta,\mu,\epsilon_{i})-(\epsilon_{i}-\mu)\omega^{2}
(\beta,\mu,\epsilon_{i}).
\end{equation}
Its solution has the same usual form of Fermi-Dirac statistics for
the free electron gas as follows,
\begin{equation}
\omega(\beta,\mu,\epsilon_{i})=\frac{1}{1+e^{\beta(\epsilon_{i}-\mu)}},
\end{equation}
with energy levels $\{\epsilon_{i}\}$ determined by Eq. (29).

\vspace{0.35in}
\section{Discussions and Conclusions}

In this paper, we have solved the set of hierarchy Bloch equations
for the reduced statistical density operator in a grand canonical
ensemble under single-orbital approximation for the identical
fermion system without or with two-body interaction. For the case
without any interaction, we not only get a plane wave solution for
the eigenstates and eigenvalues, but also recover the usual
Fermi-Dirac distribution of a free electron gas. For the situation
with two-body interaction, we obtain an eigen-equation for the
single-particle states. It is the extension of usual commonly used
Hartree-Fock equation at the absolute zero temperature to the case
of any finite temperature. The average occupation number formula
for each single-particle state is also obtained, which has the
same analytical form as that for the free electron gas with the
single-particle state energy determined by the Hartree-Fock
equation at finite temperature (29).

From Eqs. (21), (22) and (30), we see that the Coulomb operator
$J$, the exchange operator $K$, and therefore the Fock operator
$F$ are both the coherent and the incoherent superpositions of
single-particle states. They are all temperature-dependent through
an incoherent superposition factor, the Fermi-Dirac distribution,
$\omega(\beta,\mu,\epsilon_{i})$. Therefore, the mean force and
corresponding microscopic structure are temperature-dependent.

It is easy to see that, when temperature is zero, only $N$
single-particle states with energy levels small than the chemical
potential $\mu$ are occupied as for the free electron gas. That
is, the fact that only lowest $N$ orbitals (or holes) are occupied
holds also for the interacting case. Therefore, only a single
determinant wavefunction is enough for a description of
interacting $N$-particle wavefunctions. Say in another way, there
exists a correspondingly effective and strict one-particle state
description of an interacting $N$-particle system. This is also
the spirit of current density functional theory [17], and is
consistent with the third law of thermodynamics.

Even though we have simplified the structure issue for a molecule
or a solid at finite temperature, the equations we obtained in
this paper will find wide applications. The examples include the
investigation of temperature effects on the $\it{average}$
structure and the electronic spectra of macromolecules and the
study of conduction electrons in metals and so forth [18-21].

\end{document}